\documentclass[aps,amsmath,showpacs,amsfonts,10pt]{revtex4}
\usepackage{epsfig,graphicx}
\usepackage[english]{babel}
\usepackage{amsfonts}
\usepackage{amsmath}
\usepackage{latexsym}
\usepackage{graphics,bm}
\usepackage{subfigure}
\usepackage{dcolumn}
\usepackage{bm}
\usepackage{rotating}

\begin{document}

\title{Superfluid properties of a Bose-Einstein condensate in an optical lattice confined in a cavity}

\author{Aranya B Bhattacherjee}
\affiliation{Max Planck-Institute f\"ur Physik komplexer Systeme, N\"othnitzer Str.38, 01187 Dresden, Germany }

\begin{abstract}
We study the effect of a one dimensional optical lattice in a cavity field with quantum properties on the superfluid dynamics of a Bose-Einstein condensate(BEC). In the cavity the influence of atomic backaction and the external driving pump become important and modify the optical potential. Due to the coupling between the condensate wavefunction and the cavity modes, the cavity light field develops a band structure. This study reveals that the pump and the cavity emerges as a new handle to control the superfluid properties of the BEC.
\end{abstract}


\maketitle

\section{Introduction}

Cold atoms in optical lattices exhibit phenomena typical of solid state physics like the formation of energy bands, Josephson effects and Bloch oscillations. Many of these phenomena have been already the object of experimental investigations. For a recent review see \cite{Morsch06}. As the light fields that are used to create the optical lattices are intense and strongly detuned from any atomic transition, their properties can be safely approximated by classical fields. However, if the system is confined in a high-Q cavity, the quantum properties of the field becomes important, and the atoms move in quantized potentials.

Experimental implementation of a combination of cold atoms and cavity QED (quantum electrodynamics) has made significant progress \cite{Nagorny03,Sauer04,Anton05}. Theoretically there have  been some interesting work on the correlated atom-field dynamics in a cavity. It has been shown that the strong coupling of the condensed atoms to the cavity mode changes the resonance frequency of the cavity \cite{Horak00}. Finite cavity response times lead to damping of the coupled atom-field excitations \cite{Horak01}. The driving field in the cavity can significantly enhance the localization and the cooling properties of the system\cite{Griessner04,Maschler04}. It has been shown that in a cavity the atomic back action on the field introduces atom-field entanglement which modifies the associated quantum phase transition \cite{Maschler05,Mekhov07}. The light field and the atoms become strongly entangled if the latter are in a superfluid state, in which case the photon statistics typically exhibits complicated multimodal structures \cite{Chen07}. 

In this paper, we will investigate some structural properties of the coupled condensate (these are large atom samples)-light field system in the optical lattice potential of the cavity. In contrast to ref.\cite{Maschler05}, we treat the atoms within the mean-field framework. This is a valid approximation, if the number of atoms is large. The mean-field approach allows one to determine the Bloch dynamics and the Bogoliubov excitations in a straight forward manner. The long range atom-atom interactions are known to persist even in the bad cavity limit (which we assume here)and in the semiclassical limit \cite{Maschler05}. On the other hand, treating the atoms quantum mechanically, the correlation between the photon-number fluctuations and the atom-number fluctuations becomes important and will be reflected in measurable frequency dependent observables such as cavity transmission spectra \cite{Mekhov07}. As we are interested in the limit where the condensate is not destroyed by the light field, we will assume a large detuning of the pump field from the atomic resonance. Morever as mentioned earlier, we will be in the bad cavity limit where the cavity decay dominate over the spontaneous decay of all atoms thus allowing us to omit the effect of atomic decay. We also assume that the induced resonance frequency shift of the cavity is much smaller than the longitudinal mode spacing, so that we restrict the model to a single longitudinal mode. An important feature produced by the periodic potential is the occurrence of a typical band structure of the condensate in the energy spectra. Because of the strong coupling of the condensate wave function to the cavity modes, a band structure of the condensate also leads to a band structure of the intracavity light fields. This in turn influences the Bloch energies, effective mass, Bogoliubov excitations and the superfluid fraction of the BEC. We will explicitly discuss the change in the behavior of the system as a function of the quantum optical lattice depth, the pump-cavity detuning and the pump amplitude. This study reveals that the pump and the cavity now emerges as a new handle to control the coherence properties of the BEC, which offer the potential for improved interferometric technique, quantum information processing and efficient control of nonlinear excitations such as solitons. A wealth of new phenomena can be expected in the many-body physics of quantum gases with pump-cavity mediated interaction. Expressions for the tunneling parameter, the Bloch energy, the Bogoliubov spectrum and the effective mass in a quantum optical lattice are new results, derived here for the first time.

\section{Optical lattice in a cavity}

In this section, we introduce our model and calculate the effective optical lattice for which we follow earlier works\cite{Horak00,Horak01,Griessner04,Maschler04,Maschler05}. We consider an elongated cigar shaped Bose-Einstein condensate of $N$ two-level atoms with mass $m$ and transition frequency $\omega_{a}$ strongly interacting with a single one-dimensional standing wave quantized single cavity mode of frequency $\omega_{c}$. In order to create an elongated BEC, the frequency of the harmonic trap along the transverse direction should be much larger than one in the axial (along the direction of the optical lattice) direction. The system is also coherently driven by a laser field with frequency $\omega_{p}$ through the cavity mirror with amplitude $\eta$. Both the standing wave and the pump laser are weak(small photon number) so that we treat them quantum-mechanically. The quantum nature of the light is not an essential requirement for the results that follow but is essential for measurement of the transmission spectra which depends on the photon number $<\hat a^{\dagger} \hat a>$ \cite{Mekhov07}. It is well known that high-Q optical cavities can significantly isolate the system from its environment, thus strongly reducing decoherence and ensuring that the light field remains quantum-mechanical for the duration of the experiment. The quantum nature of the light field can be used to probe the number-statistics of the matter-wave field due to atom-photon entanglement. Classical radiation only provides information on some average of the atomic occupation numbers. The harmonic confinement along the directions perpendicular to the optical lattice is taken to be large so that the system effectively reduces to one-dimension. This system is modeled by a Jaynes-Cummings type of Hamiltonian $(H_{JC})$ in a rotating wave and dipole approximation \cite{Maschler05}

\begin{figure}[t]
\hspace{-2.0cm}
\includegraphics{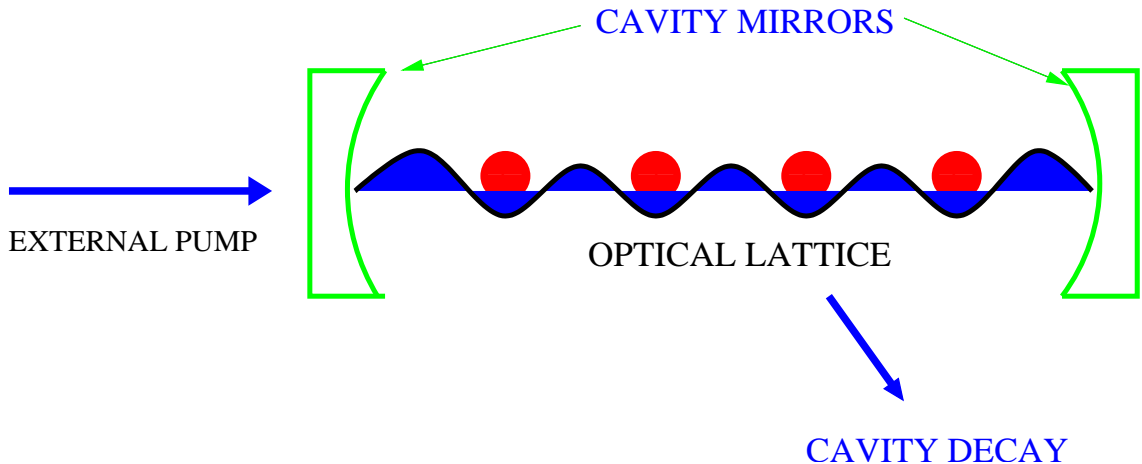} 
\caption{Schematic representation of the ultracold atoms trapped in an optical lattice confined in a cavity. An external pump is also incident from one of the side mirrors}.
\label{1}
\end{figure}

\begin{equation}
H_{JC}=\dfrac{p^2}{2m}-\hbar \Delta_{a} \sigma^{+} \sigma^{-} -\hbar \Delta_{c}\hat{a}^{\dagger}\hat{a}-i\hbar g(x)\left[ \sigma^{+}\hat{a}-\sigma^{-}\hat{a}^{\dagger}\right]-i\eta(\hat{a}-\hat{a}^{\dagger}), 
\end{equation}

where $\Delta_{a}=\omega_{p}-\omega_{a}$ and $\Delta_{c}=\omega_{p}-\omega_{c}$ are the large atom-pump and cavity-pump detuning, respectively and  $\Delta_{c}>\Delta_{a}$. In this work we will consider only the case  $\Delta_{a}>0$. Here $\sigma^{+} , \sigma^{-}$ are the Pauli matrices and $\eta$ is the strength of the external pump. The atom-field coupling is written as $g(x)=g_{0} \cos(kx)$. Here $\hat{a}$ is the annihilation operator for a cavity photon. Since the detuning $\Delta_{a}$ is large, spontaneous emission is negligible and we can adiabatically eliminate the excited state using the Heisenberg equation of motion $\dot{\sigma^{-}}=\dfrac{i}{\hbar}\left[ H_{JC},\sigma^{-}\right] $. This yields the single particle Hamiltonian

\begin{equation}
H_{0}=\dfrac{p^2}{2m}-\hbar \Delta_{c}\hat{a}^{\dagger}\hat{a}+\dfrac{\hbar U_{0} \cos^2(kx)}{\Delta_{a}}\left[ 1+\hat{a}^{\dagger} \hat{a}\right]-i\eta(\hat{a}-\hat{a}^{\dagger}). 
\end{equation}

The parameter $U_{0}=\dfrac{g_{0}^{2}}{\Delta_{a}}$ is the optical lattice barrier height per photon and represents the atomic backaction on the field \cite{Maschler05}. Here we will always take $U_{0}>0$. In this case the condensate is attracted to the nodes of the light field and hence the lowest bound state is localized at these positions which leads to a reduced coupling of the condensate to the cavity compared to that for $U_{0}<0$.  Along $x$, the cavity field forms an optical lattice potential of period $\lambda/2$ and depth $\hbar U_{0}(\hat{a}^{\dagger}\hat{a}+1)$. We now write the Hamiltonian in a second quantized form including the two body interaction term.

\begin{equation}
H=\int d^3 x \Psi^{\dagger}(\vec{r})H_{0}\Psi(\vec{r})+\dfrac{1}{2}\dfrac{4\pi a_{s}\hbar^{2}}{m}\int d^3 x \Psi^{\dagger}(\vec{r})\Psi^{\dagger}(\vec{r})\Psi(\vec{r})\Psi(\vec{r}),
\end{equation}

where $\Psi(\vec{r})$ is the field operator for the atoms. Here $a_{s}$ is the two body $s$-wave scattering length. The corresponding Bose-Hubbard Hamiltonian can be derived by writing $\Psi(\vec{r})=\sum_{i} \hat{b}_{i} w(\vec{r}-\vec{r}_{i})$, where $w(\vec{r}-\vec{r}_{i})$ is the Wannier function and $\hat{b}_{i}$ is the corresponding annihilation operator for the bosonic atom. Retaining only the lowest band with nearest neighbor interaction, we have

\begin{eqnarray}
H = E_{0}\sum_{j}\hat{b}_{j}^{\dagger}\hat{b}_{j}&+&E\sum_{j}\left(\hat{b}_{j+1}^{\dagger}\hat{b}_{j}+\hat{b}_{j+1}\hat{b}_{j}^{\dagger} \right)+\hbar U_{0}(\hat{a}^{\dagger}\hat{a}+1)\left\lbrace J_{0}\sum_{j}\hat{b}_{j}^{\dagger}\hat{b}_{j}+J \sum_{j}\left(\hat{b}_{j+1}^{\dagger}\hat{b}_{j}+\hat{b}_{j+1}\hat{b}_{j}^{\dagger} \right)\right\rbrace\nonumber \\&-&\hbar \Delta_{c} \hat{a}^{\dagger}\hat{a}-i\hbar \eta (\hat{a}-\hat{a}^{\dagger})+\dfrac{U}{2}\sum_{j}\hat{b}_{j}^{\dagger}\hat{b}_{j}^{\dagger}\hat{b}_{j}\hat{b}_{j}\; 
\end{eqnarray}

where

\begin{eqnarray}
U&=&\dfrac{4\pi a_{s}\hbar^{2}}{m}\int d^3 x|w(\vec{r})|^{4}\nonumber \\
E_{0}&=&\int d^3 x w(\vec{r}-\vec{r}_{j})\left( -\dfrac{\hbar^2 \nabla^{2}}{2m}\right)w(\vec{r}-\vec{r}_{j})\nonumber \\
E &=&\int d^3 x w(\vec{r}-\vec{r}_{j})\left( -\dfrac{\hbar^2 \nabla^{2}}{2m}\right)w(\vec{r}-\vec{r}_{j \pm 1})\nonumber \\
J_{0}&=&\int d^3 x w(\vec{r}-\vec{r}_{j}) \cos^2(kx)w(\vec{r}-\vec{r}_{j})\nonumber \\
J &=&\int d^3 x w(\vec{r}-\vec{r}_{j}) \cos^2(kx)w(\vec{r}-\vec{r}_{j \pm 1}).
\end{eqnarray}

The nearest neighbor nonlinear interaction terms are usually very small compared to the onsite interaction and are neglected as usual. The onsite energies $J_{0}$ and $E_{0}$ are set to zero. We now write down the Heisenberg equation of motion for the cavity photons $\hat{a}$ and the bosonic field operator $\hat{b}$ as

\begin{equation}
\dot{\hat{a}}=-iU_{0}\left\lbrace J_{0}\sum_{j}\hat{b}_{j}^{\dagger}\hat{b}_{j}+J \left(\hat{b}_{j+1}^{\dagger}\hat{b}_{j}+\hat{b}_{j+1}\hat{b}_{j}^{\dagger} \right)\right\rbrace \hat{a}+\eta+i\Delta_{c} \hat{a}-\gamma \hat{a}
\end{equation}

\begin{equation}
\dot{\hat{b}}_{j}=-iU_{0}\left( 1+\hat{a}^{\dagger} \hat{a} \right)J\left\lbrace \hat{b}_{j+1}+\hat{b}_{j-1} \right\rbrace-\dfrac{iE}{\hbar}\left\lbrace\hat{b}_{j+1}+\hat{b}_{j-1}  \right\rbrace - \dfrac{iUn_{0}}{\hbar}\hat{b}_{j}.
\end{equation}

Here we have introduced $\gamma$ as the field damping rate. Equation (6) for the cavity field and equation (7) for the atomic field represents a set of coupled equations describing the dynamics of the compound system formed by the condensate and the optical cavity. We will work in the bad cavity limit, where typically, $\gamma$ is the fastest time scale ( this means that the cavity decay rate is much larger than the oscillation frequency of bound atoms in the optical lattice of the cavity ). In this limit the intracavity field adiabatically follows the condensate wavefunction, and hence we can put $\dot{\hat{a}}=0$. We treat the BEC within the mean field framework and assume the tight binding approximation where we replace $\hat{b}_{j}$ by $\phi_{j}$ and look for solutions in the form of Bloch waves 

\begin{equation}
\phi_{j}=u_{k}exp(ikjd)exp(-i\mu t/ \hbar).
\end{equation}

Here $\mu$ is the chemical potential, $d$ is the periodicity of the lattice and $\sum_{j}\hat{b}_{j}^{\dagger} \hat{b}_{j}=|u_{k}|^{2}=n_{0}$ (atomic number density). The tight binding approximation becomes more and more accurate as the depth of the optical lattice increases and this can be achieved by a strong laser which makes up the optical lattice or a strong external pump. Alternatively, we can also achieve the tight binding regime by keeping the pump-cavity detuning $\Delta_{c}\approx 2Jn_{0}U_{0} \cos(kd)$. This yields from equation (6)

\begin{equation}
\hat{a}=\dfrac{\eta}{\left\lbrace \gamma-i\left( \Delta_{c}-2Jn_{0}U_{0}\cos(kd)\right) \right\rbrace }.
\end{equation}

Interestingly we find that due to the atomic backaction, the quantum state of the cavity field varies along the Brillioun zone. The cavity photons develops a band structure due to the strong coupling with the condensate, analogous to photonic band gap materials. The concept of photonic band gaps in optical lattices has been known for quite some time \cite{Deutsch94}. This expression shows that the cavity photons are created by the scattering through the atoms which are coherently driven by the external pump. Interestingly, the average photon number $<\hat a^{\dagger} \hat a>$ measures the light transmission spectra and is different for the Mott insulator (MI) and the superfluid phase (SF) \cite{Mekhov07}. The atomic backaction then also modifies the effective optical lattice potential as $V_{op}=\hbar U_{0}(1+\hat{a}^{\dagger} \hat{a})$

\begin{equation}
V_{op}=\hbar U_{0}\left(1+\dfrac{\eta^{2}}{\gamma^{2}+\left[ \Delta_{c}-2Jn_{0}U_{0}\cos(kd)\right]^2 } \right). 
\end{equation}

\begin{figure}[t]
\hspace{-1.5cm}
\includegraphics{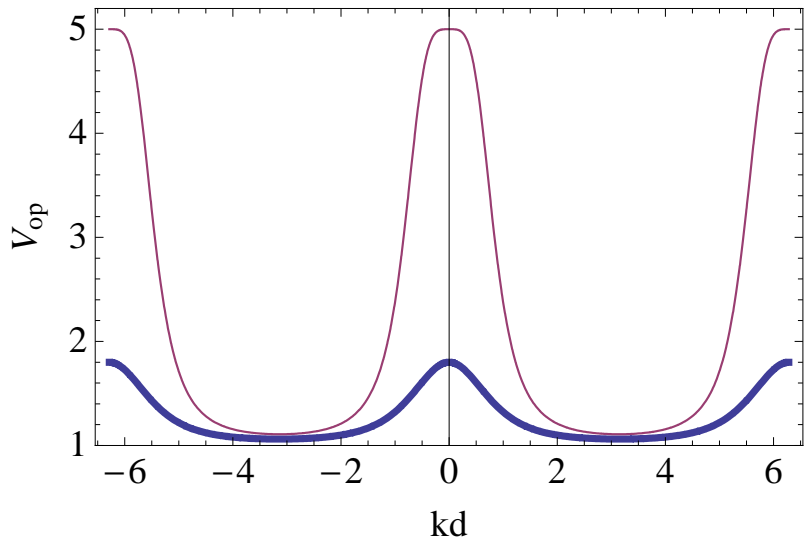} 
\caption{The effective optical lattice potential modified by the atomic backaction as a function of the quasi-momentum for two different values of the pump-cavity detuning $\frac{\Delta_{c}}{\gamma}=5$ (bold line), $\frac{\Delta_{c}}{\gamma}=3$ (thin line). The pump amplitude is $\frac{\eta}{\gamma}=2$ and the optical lattice barrier height per photon is $\frac{U_{0}}{\gamma}=1$ and $\frac{JU_{0}n_{0}}{\gamma}=1.5$ . A larger pump-cavity detuning reduces the effective optical potential height.}
\label{fig:figure_2}
\end{figure}

As the condensate moves across the Brillioun zone, the atom-field interaction changes and as a result the optical lattice is also continuously modified. A decrease in the pump-cavity detuning $\Delta_{c}$ increases this atomic backaction and the potential increases. This is evident from the figure 2 where we have plotted the effective potential as a function of the quasi-momentum for two different values of pump-cavity detuning, $\frac{\Delta_{c}}{\gamma}=5$ (bold line),$\frac{\Delta_{c}}{\gamma}=3$ (thin line). The barrier height is maximum when $\Delta_{c}=2Jn_{0}U_{0} \cos(kd)$. As evident from eqn.(10), the effective optical lattice increases with the pump amplitude. In the absence of the pump the $k$ dependence vanishes.

\section{Bloch energy and Effective mass}

In this section we will calculate the Bloch chemical potential, the lowest Bloch band and the corresponding effective mass. We again assume mean-field solutions and substituting equation (8) into equation (7) and also using equation (9), we get the chemical potential of the system as

\begin{equation}
\mu=Un_{0}-2J_{eff}(k) \cos(kd),
\end{equation}

where we define an effective tunneling parameter which also has a $k$ dependence as
 
\begin{equation}
J_{eff}(k)=-E-\hbar U_{0}J\left(1+\dfrac{\eta^{2}}{ \gamma^{2}+\left[ \Delta_{c}-2Jn_{0}U_{0}\cos(kd)\right]^2 } \right). 
\end{equation}

Note that the overall sign of $J_{eff}(k)$ must be positive since $E$ which is determined by the kinetic energy term has a negative sign and its magnitude is larger than the second term. The expression for $J_{eff}(k)$ implies that as the pump amplitude increases, the tunneling between neighbouring wells decreases since the height of the barriers increases. For a fixed pump amplitude, tunneling is minimized when $\Delta_{c}=2Jn_{0}U_{0} \cos(kd)$. The energy per particle is defined as

\begin{equation}
\epsilon(k)=\dfrac{1}{n_{0}}\int \mu \, dn_{0}.
\end{equation}

This yields

\begin{equation}
\epsilon(k)=\dfrac{U n_{0}}{2}+2(E+\hbar U_{0}J)\cos(kd)-\dfrac{\hbar \eta^{2}}{\gamma n_{0}^2}tan^{-1}\left[\dfrac{\Delta_{c}-2U_{0}n_{0}J \cos(kd)}{\gamma} \right]. 
\end{equation}

\begin{figure}[t]
\hspace{-1.5cm}
\includegraphics{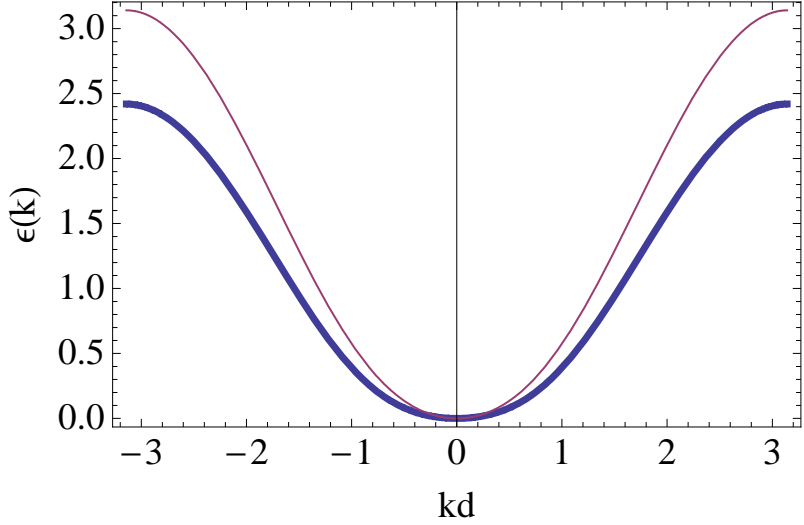} 
\caption{The Bloch spectrum in the units of recoil energy for pump amplitude $\frac{\eta}{\gamma}=2$, $\frac{E}{\gamma}=-2$ (kinetic energy contribution to the tunneling), $\frac{\Delta_{c}}{\gamma}=5$ (pump-cavity detuning) and for two values of the potential energy contribution to the tunneling parameter $\frac{JU_{0}}{\gamma}=1.0$ (thin line), $\frac{JU_{0}}{\gamma}=1.2$ (bold line). The ground state energy has been subtracted. The Bloch energy is reduced in a deeper quantum optical lattice. $\frac{JU_{0}}{\gamma}$ can also be termed as effective optical lattice barrier height and can be varied by changing the pump-atom detuning or the laser intensity of the optical lattice. }  
\label{fig:figure_3}
\end{figure}

\begin{figure}[t]
\hspace{-1.5cm}
\includegraphics{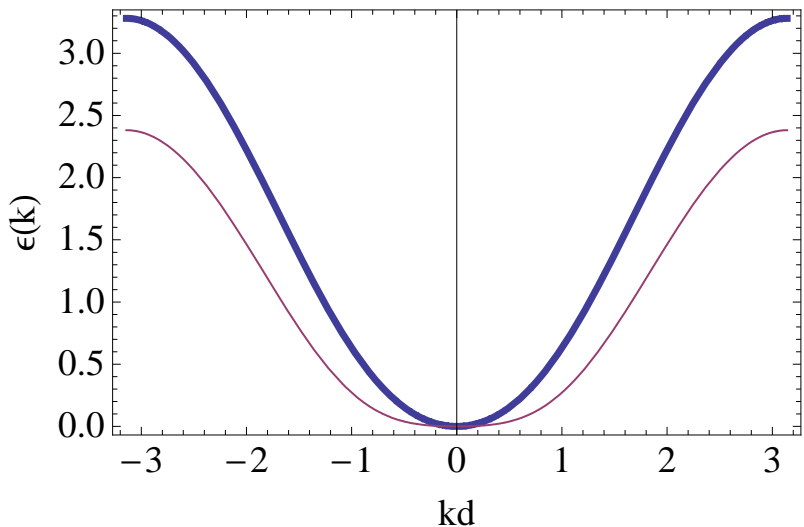} 
\caption{The Bloch energy in the units of recoil energy for $\frac{JU_{0}}{\gamma}=1$ (effective optical lattice height), $\frac{E}{\gamma}=-2$, $\frac{\Delta_{c}}{\gamma}=5$ (pump-cavity detuning) and for two two values of the pump amplitude, $\frac{\eta}{\gamma}=3.0$ (thin line),  $\frac{\eta}{\gamma}=2.0$ (bold line). The ground state energy has been subtracted. A stronger pump reduces the Bloch energy because the corresponding effective optical lattice also becomes deeper.}
\label{fig:figure_4}
\end{figure}

The second term in equation (14) is the tight binding expression for the energy of a Bloch state for a single particle in an optical lattice while the third term is the influence of the cavity photons on the Bloch energy induced by the external pump. The pump plays a very important and interesting role in manipulating the superfluid properties of the system.  The energy per particle $\epsilon(k)$ of stationary Bloch configuration consists of the motion of the whole condensate and carries current, constant in time and uniform in space(Bloch bands). Figure 3 shows a plot of the Bloch energy of equation (14) as a function of $kd$ for $\dfrac{JU_{0}}{\gamma}=1.0$ (thin line)$\dfrac{JU_{0}}{\gamma}=1.2$ (thick line), $\dfrac{\eta}{\gamma}=2.0$, $\dfrac{E}{\gamma}=-2.0$ and $\dfrac{\Delta_{c}}{\gamma}=-5.0$. The ground state energy has been subtracted in figure 3. The Bloch spectrum is found to be suppressed at higher value of the optical lattice depth per photon. The effect is more pronounced near the band edge.In a similar manner we study the influence of the pump on the Bloch spectrum in figure 4. A stronger pump reduces the Bloch energy because the corresponding effective optical lattice barrier height also increases. In the absence of the pump, the expression for the Bloch energy (equation 14) reduces to the usual expression found for BEC in an optical lattice in the absence of cavity \cite{Menotti03}. The Bloch energy derived here for a BEC in an optical lattice confined in a cavity has a behaviour similar to that found for an optical lattice without a cavity \cite{Menotti03} but the important point is that the cavity and the pump parameters now emerge as a new tool to manipulate properties of the BEC. We already see that the expression for the Bloch energy (equation 14) is modified by the cavity. Let us now derive the effective mass by studying the low-k behaviour of the lowest band $\epsilon (k)$. The effective mass of the atoms in the optical lattice is defined as $\dfrac{1}{m^{*}}=\dfrac{1}{\hbar^{2}}\dfrac{\partial^{2} \epsilon(k)}{\partial k^{2}}|_{k=0}$, this yields

\begin{equation}
m^{*}=\dfrac{-\hbar^{2}}{2d^2(E+\hbar U_{0} J)}\left\lbrace\dfrac{n_{0}\gamma^{2}\left[ 1+\left( \dfrac{\Delta_{c}}{\gamma}-\dfrac{2n_{0}U_{0}J}{\gamma}\right)^{2} \right] }{n_{0}\gamma^{2}\left[ 1+\left( \dfrac{\Delta_{c}}{\gamma}-\dfrac{2n_{0}U_{0}J}{\gamma}\right)^{2}\right] +\dfrac{\hbar U_{0}J \eta^{2}}{(E+\hbar U_{0}J)}} \right\rbrace.
\end{equation}

A plot of the ratio of the effective mass to the bare mass as a function of $\dfrac{JU_{0}}{\gamma}$ for two different values of the pump amplitude is depicted in figure 5. For $\dfrac{JU_{0}}{\gamma}\rightarrow 0$ (as the optical lattice height vanishes), the effective mass tends to the bare mass $m$. For small values of the lattice depth, the effective mass increases slowly and the two curves, plotted for two different values of the pump amplitudes are not distinguishable. For large $\dfrac{JU_{0}}{\gamma}$ however, the effective mass increases strongly due to the decrease of the tunneling between neighboring wells and the increase is faster for a stronger pump. The effect of the pump is to increase the value of $m^*$ as a consequence of an increase in the effective optical potential. As we increase the pump, the optical lattice barrier height increases and the tunneling decreases accompanied by an increase in the effective mass. The ground state wave function becomes effectively more localized for stronger pump and larger optical potentials. This result is in accordance with earlier work \cite{Horak01}. The sound velocity is found as $\sqrt{\dfrac{Un_{0}}{m^*}}$ and it decreases as the effective optical lattice in the cavity is made deeper. The influence of cavity-pump detuning on the effective mass is illustrated in figure 6. Interestingly, for $\omega_{p}>\omega_{c}$ (positive detuning), there is a sharp increase in the effective mass and it shows a maxima at $\Delta_{c}=2n_{0}U_{0}J$. For this value of $\Delta_{c}$, the lattice depth is maximum and the superfluid component is expected to be minimum. Hence the detuning gives a handle to control the superfluid properties and nonlinear excitations such as solitons in optical cavities. Properties of nonlinear excitations depends on the effective mass.  The role of the interactions is to decrease the value of the effective mass as a consequence of the broadening of the wavefunction caused by the repulsion, which favours tunneling, contrasting the effect of the pump, which favours localization due to an increase in the optical lattice depth. In general there are two effective masses for a BEC in an optical lattice\cite{Menotti03}. In BEC, because of the nonlinearity, the two relevant energies, the Bloch energy and the chemical potential has different curvatures. This gives rise to two different effective masses. 

\begin{figure}[t]
\hspace{-1.5cm}
\includegraphics{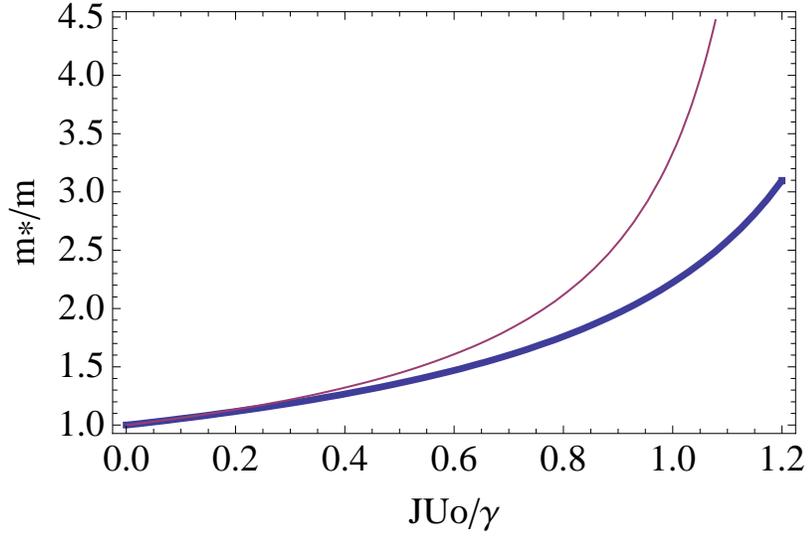} 
\caption{A plot of the effective mass as a function of the effective optical lattice barrier height $\frac{JU_{0}}{\gamma}$ for two different values of the pump amplitude,$\frac{\eta}{\gamma}=2$ (thin line), $\frac{\eta}{\gamma}=1$ (bold line). Other parameters are $\frac{E}{\gamma}=-2$ (kinetic energy contribution to the tunneling), $\frac{\Delta_{c}}{\gamma}=5$ (pump-cavity detuning). The effect of increasing the pump amplitude is to increase the effective mass.For $\dfrac{JU_{0}}{\gamma}\rightarrow 0$ (as the optical lattice height vanishes), the effective mass tends to the bare mass $m$.}
\label{fig:figure_5}
\end{figure}

\begin{figure}[t]
\hspace{-1.5cm}
\includegraphics{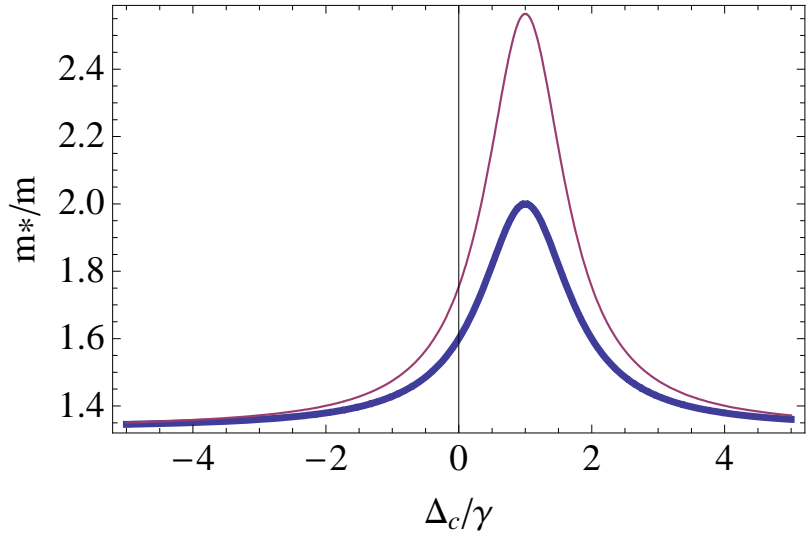} 
\caption{A plot of the effective mass as a function of the pump-cavity detuning $\frac{\Delta_{c}}{\gamma}$ for two different values of the pump amplitude, $\frac{\eta}{\gamma}=2$ (thin line), $\frac{\eta}{\gamma}=1$ (bold line). Other parameters are $\frac{E}{\gamma}=-2$, $\frac{Jn_{0}U_{0}}{\gamma}=0.5$. The effective mass shows a maxima at $\Delta_{c}=2n_{0}U_{0}J$}
\label{fig:figure_6}
\end{figure}

\section{Bogoliubov Dispersion Relation}

In this section we study the spectrum of elementary excitations. We will focus our attention on the excitations relative to the ground state ($k=0$). The Bogoliubov spectrum of elementary excitation describes the energy of small perturbations with quasi-momentum $q$ on top of a macroscopically populated state with quasi-momentum $k$. The ground state solution ($k=0$) for the translationally invariant lattice gives the eigenvalue: 

\begin{equation}
\mu = n_{0} U - 2J_{eff}^{0},
\end{equation}

where $J^{0}_{eff}=J_{eff}(k=0)$. We now include the quantum fluctuations in our
description of the system using the Bogoliubov approximation, where we
suppose that we can write the full annihilation operator in terms of the $c$ -number part ($\phi_{j}$) and a fluctuation operator ($\hat{\delta}_{j}$) thus: 

\begin{equation}
\hat{b}_{j} = (\phi_{j}+\hat{\delta}_{j}) e^{-\mathrm{i}\frac{\mu t}{\hbar}} \;.
\end{equation}
This form is useful when we are looking at the properties of a
time-independent or adiabatic ground state. In using this method we are
assuming that the fluctuation part is small. As mentioned earlier the optical lattice potential in the cavity depends on the BEC wavefunction. This implies that fluctuations in the wavefunction would naturally change the optical lattice potential. Since the fluctuations are small, we will neglect fluctuation induced changes in the lattice potential. The Bogoliubov equations for the lattice have the following form:

\begin{equation}
\mathrm{i}\hbar\, \partial_{t}\hat{\delta}_{j} = (2n_{0}U-\mu) \hat{\delta}%
_{j} - J_{eff}^{0}(\hat{\delta}_{j+1} + \hat{\delta}_{j-1}) + n_{0}U \hat{\delta}%
_{j}^{\dagger} \;.
\end{equation}
This is solved by constructing quasi-particles for the lattice which
diagonalize the Hamiltonian \cite{Ana03}, i.e 

\begin{eqnarray}
\hat{\delta}_{j} &=& \frac{1}{\sqrt{I}} \sum_{q} [u^{q}\hat{\alpha}_{q}\; e^{%
\mathrm{i}(qid-\omega _{q}t)} - v^{q
\ast}\hat{\alpha}_{q}^{\dagger}\; e^{-\mathrm{i}(qid-\omega
_{q}t)}]
\label{qua1} \\
\hat{\delta}_{j}^{\dagger} &=& \frac{1}{\sqrt{I}} \sum_{q} [u^{q \ast }\hat{%
\alpha}_{q}^{\dagger}\;e^{-\mathrm{i}(qid-\omega _{q}t)} - v^{q}\hat{\alpha}_{q}\; e^{%
\mathrm{i}(qid-\omega _{q}t)}] \;,  \label{qua2}
\end{eqnarray}
where $d$ is the lattice spacing. The quasi-particle operators obey the
usual Bose commutation relations: 
\begin{equation}
\big[ \hat{\alpha}_{q},\hat{\alpha }_{q^{\prime}}^{\dagger}\big] =
\delta_{qq^{\prime}}
\end{equation}

We then find the following equations for the excitation amplitudes and
frequencies, 
\begin{eqnarray}
\hbar \omega_{q} u^{q} &=& \Big[ n_{0}U + 4J_{eff}^{0} \sin^{2} \Big(\frac{qd}{2}\Big) %
\Big] u^{q} - n_{0} U v^{q}, \\
-\hbar \omega_{q} v^{q} &=& \Big[ n_{0} U + 4J_{eff}^{0} \sin^{2}\Big(\frac{qd}{2}\Big)%
\Big] v^{q} - n_{0} U u^{q}.
\end{eqnarray}
Thus, the expressions for the $u^{q}$ and $v^{q}$ yield: 
\begin{eqnarray}
|u^{q}|^{2} &=& \frac{K(q)+n_{0}U+\hbar \omega _{q}}{2\hbar \omega_{q}} \\
|v^{q}|^{2} &=& \frac{K(q)+n_{0}U-\hbar \omega _{q}}{2\hbar \omega_{q}} \;,
\end{eqnarray}
where the phonon excitation frequencies are given by: 
\begin{eqnarray}
\hbar \omega _{q} &=& \sqrt{K(q)[2n_{0}U+K(q)]} \\
K(q) &=& 4J_{eff}^{0} \sin^{2}\Big(\frac{qd}{2}\Big) \;.  \label{kq}
\end{eqnarray}

\begin{figure}[t]
\hspace{-1.5cm}
\includegraphics{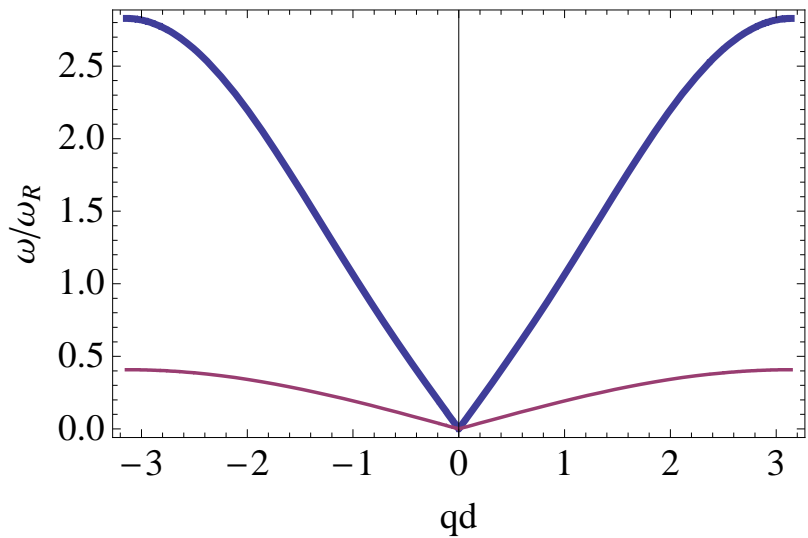} 
\caption{Bogoliubov spectrum (as a function of quasi-momentum) in the units of recoil frequency for effective optical lattice barrier height $\frac{JU_{0}}{\gamma}=1$, $\frac{E}{\gamma}=-2$ (kinetic energy contribution to tunneling), $\frac{U}{\gamma}=2$ (two-body interaction strength), $\frac{\Delta_{c}}{\gamma}=3$ (pump-cavity detuning) and two values of the pump amplitude $\frac{\eta}{\gamma}=1.4$ (thin line), $\frac{\eta}{\gamma}=1.0$ (thick line). The Bogoliubov energy decreases and the band becomes flatter for a stronger pump.}
\label{fig:figure_7}
\end{figure}

\begin{figure}[t]
\hspace{-1.5cm}
\includegraphics{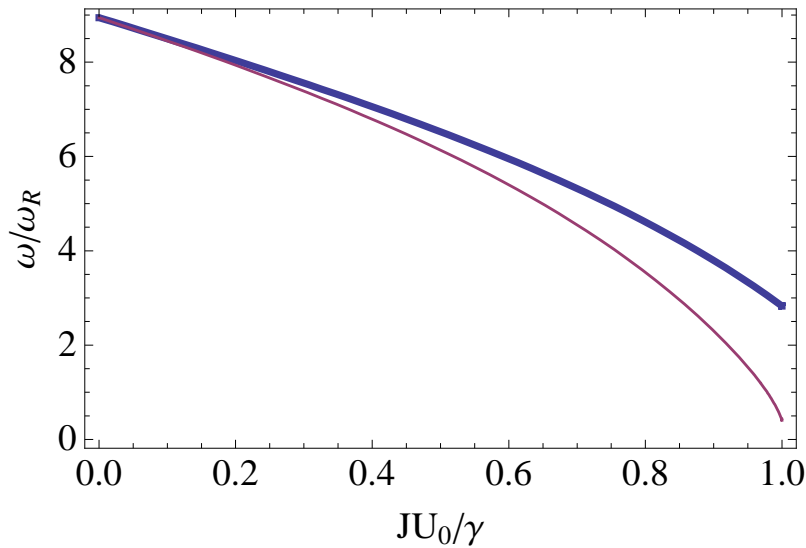} 
\caption{Bogoliubov excitation energy in the units of recoil frequency as a function of the effective optical lattice height $\frac{JU_{0}}{\gamma}$ for $\frac{E}{\gamma}=-2$,  $\frac{U}{\gamma}=2$ (two-body interaction strength),  $\frac{\Delta_{c}}{\gamma}=3$ (pump-cavity detuning), $\frac{\eta}{\gamma}=1.4$ (thin line), $\frac{\eta}{\gamma}=1.0$ (thick line), $qd=\pi$ (edge of the zone). The Bogoliubov excitation energy decreases with increasing strength of the optical lattice (or decreasing $\Delta_{c}$).}
\label{fig:figure_8}
\end{figure}

The effect of the optical cavity and the external pump on the Bogoliubov spectrum comes in through the tunneling parameter $J_{eff}^{0}$. Figure 7 shows the Bogoliubov spectrum for two different pump amplitudes. The pump is found to suppress the Bogoliubov spectrum which is due to a corresponding increase in the effective mass. At $\dfrac{\eta}{\gamma}=1.0$ (thick line), the curve resembles the dispersion in the uniform case (absence of optical lattice): both the phononic linear regime and the quadratic regime are visible. When the potential is made deeper by increasing the pump $\dfrac{\eta}{\gamma}=1.4$, the band becomes flatter (thin line). As a consequence, the slope of the phononic regime decreases. This also reflects the behavior of the velocity of sound.In a recent experiment on the cavity QED with a BEC \cite{Ferd07}, it was observed that the energy of a single excitation increases with the increase in the cavity-atom detuning on the positive side. In other words this implies that the single excitation energy increases with decreasing strength of the optical lattice. In figure 8, we plot the Bogoliubov excitation energy at the edge of the zone ($qd=\pi$) as a function of the optical lattice barrier height per photon. In accordance with the above mentioned experiment, we make a similar observation. To conclude this section, we remark that in general the Bogoliubov excitations depends on the two different types of effective mass \cite{Menotti03}. The effects related to the difference between the two effective masses in the Bogoliubov spectrum of a condensate at rest ($k=0$) are usually negligible. In contrast, such differences becomes important when the condensate moves with a large quasi-momentum.


\section{Superfluid fraction}

In this section we calculate the superfluid fraction for our system following \cite{Ana03}. The concept of superfluidity is closely related to the existence of a condensate in the interacting many--body system. The one--body density matrix  has to have exactly one macroscopic eigenvalue which defines the number of particles in the condensate; the corresponding eigenvector describes the condensate wave function $\phi _{0}\left( \vec{r}\right) =e^{i\Theta (\vec{r})}\left| \phi _{0}\left( \vec{r}\right) \right| $. A spatially varying condensate phase, $\Theta \left( \vec{r}\right) $, is associated with a superfluid velocity field for the condensate given by

\begin{equation}
\vec{v}_{s}\left( \vec{r}\right) =\frac{\hbar }{m^*}\vec{\nabla}\Theta \left(
\vec{r}\right) .  \label{ediff}
\end{equation}

This enables us to derive an expression for the superfluid fraction, $f_{s}$. Consider a system with a finite linear dimension, $L$, in the direction of the optical lattice and a ground--state energy, $E_{0}$, calculated with periodic boundary conditions. Now we impose a linear phase variation, $\Theta \left( \vec{x}\right) =\theta x/L$ with a total twist angle $\theta $ over the length of the system in the direction of the optical lattice. The resulting ground--state energy, $E_{\theta }$ will depend on the phase twist. For very small twist angles, $\theta \ll \pi $, the energy difference, $E_{\theta }-E_{0}$, can be attributed to the kinetic energy, $T_{s}$, of the superflow generated by the phase gradient as per

\begin{equation}
E_{\theta }-E_{0}=T_{s}=\frac{1}{2}m^* Nf_{s}\vec{v}_{s}^{2},
\end{equation}

where $m^*$ is the effective mass of a single particle in the quantum optical lattice and $N$ is the total number of particles so that $mNf_{s}$ is the total mass of the superfluid component. Replacing the superfluid velocity, $\vec{v}_{s}$ with the phase gradient according to Eq.\ (\ref{ediff}) leads to the following relation for the superfluid fraction

\begin{equation}
f_{s}=\frac{2m^*}{\hbar ^{2}}\frac{L^{2}}{N}\frac{E_{\theta }-E_{0}}{\theta
^{2}}=\frac{1}{N}\frac{E_{\theta }-E_{0}}{J_{eff}^{0} \left( \Delta \theta \right) ^{2}},
\end{equation}

where the second equality applies to a lattice system on which a linear phase variation has been imposed. Here the distance between sites is $d$, the phase variation over this distance is $\Delta \theta $, and the number of sites is $I$. In this case, $J_{eff}^{0}\equiv \hbar ^{2}/(2m^* d^{2})$.

These phase factors show that the twist is equivalent to the imposition of an acceleration on the lattice for a finite time. 
We calculate the change in energy $E_{\theta}-E_{0}$ under the assumption that the phase change $\Delta \theta $ is small. 
The current operator $\hat{J}$ and the hopping operator $\hat{T}$ are given by: 

\begin{eqnarray}
\hat{J} &=&\mathrm{i}J_{eff}^{0} \sum_{i=1}^{I}(\hat{b}_{i+1}^{\dagger }\hat{b}_{i}-%
\hat{b}_{i}^{\dagger }\hat{b}_{i+1}) \\
\hat{T} &=&-J_{eff}^{0} \sum_{i=1}^{I}(\hat{b}_{i+1}^{\dagger }\hat{b}_{i}+\hat{b}%
_{i}^{\dagger }\hat{b}_{i+1})\;.
\end{eqnarray}

The change in the energy $E_{\theta }-E_{0}$ due to the imposed phase twist can now be evaluated in second order perturbation theory 

\begin{equation}
E_{\theta }-E_{0}=\Delta E^{(1)}+\Delta E^{(2)}\;.
\end{equation}

The first order contribution to the energy change is proportional to the expectation value of the hopping operator 

\begin{equation}
\Delta E^{(1)} =-\frac{1}{2}(\Delta \theta )^{2}\langle \Psi _{0}|\hat{T}|\Psi _{0}\rangle \;.
\end{equation}

Here $|\Psi _{0}\rangle $ is the ground state of the original Bose-Hubbard Hamiltonian. The second order term is related to the matrix
elements of the current operator involving the excited states $|\Psi _{\nu}\rangle $ ($\nu =1,2,...$) of the original Hamiltonian 

\begin{equation}
\Delta E^{(2)}= -(\Delta \theta
)^{2}\sum_{\nu \neq 0}\frac{|\langle \Psi _{\nu }|\hat{J}|\Psi _{0}\rangle
|^{2}}{E_{\nu }-E_{0}}\;.
\end{equation}

The superfluid fraction is then
given by the contribution of both the first and second order term. Having obtained the expressions for the excitations we can now determine the superfluid fraction in the Bogoliubov approximation. The quantity we need to calculate is just the first order term because the second order term vanishes in the Bogoliubov limit due to the translational
invariance of the lattice ,

\begin{eqnarray}
f_{\mathrm{s}} &=& \frac{1}{2N}\sum_{i=1}^{I} \langle\Psi_{0}| (\hat{\delta}%
_{i+1}^{\dagger}+\phi_{i+1})(\hat{\delta}_{i}+\phi_{i}) +(\hat{\delta}%
_{i}^{\dagger}+\phi_{i})(\hat{\delta}_{i+1}+\phi_{i+1}) |\Psi_{0}\rangle \notag \\
&=& \frac{1}{2N} \sum_{i=1}^{I} \langle\Psi_{0}| 2\phi_{i}^{2} + \hat{\delta}%
_{i+1}^{\dagger} \hat{\delta}_{i} + \hat{\delta}_{i}^{\dagger} \hat{\delta}%
_{i+1} |\Psi_{0}\rangle \;.
\end{eqnarray}
We can now express the fluctuation operators, Eqs. (\ref{qua1}) and (\ref{qua2}), in terms of the quasi-particle operators that diagonalize the quadratic Hamiltonian. This leads to the following expression for the superfluid fraction at zero temperature: 

\begin{equation}
f_{\mathrm{s}} = \frac{I}{N} \Big[\phi^{2} + \frac{1}{I} \sum_{q} |v^{q}|^{2}
\cos(qd) \Big] \;.  \label{sffrac_bog}
\end{equation}
Here the summation runs over all quasi-momenta $q = \frac{2\pi}{Id}\, j$ with $j=1,...,(I-1)$ and we have called $\phi$ the value of all $\phi_{i}$ in a translationally invariant system. This shows that in the limit of zero lattice spacing (while keeping $q$ finite) the superfluid fraction is unity as we have the normalization condition: 
\begin{equation}
I\phi^{2}+\sum_{q}|v^{q}|^{2}= N \;.
\end{equation}

\begin{figure}[t]
\hspace{-1.5cm}
\includegraphics{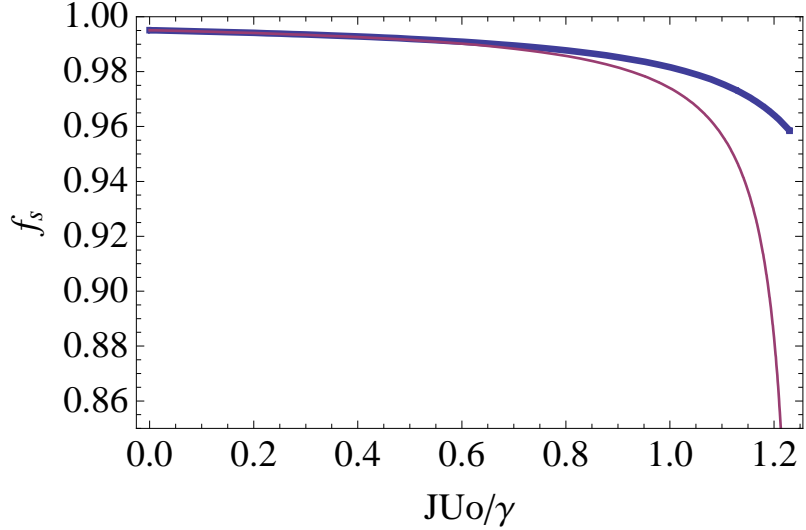} 
\caption{The superfluid fraction as a function of the effective lattice height for $\frac{\Delta_{c}}{\gamma}=4.0$ (pump-cavity detuning), $\frac{E}{\gamma}=-2.0$, $\frac{U}{\gamma}=2.0$ (two-body interaction strength) , $N=20$ and two values of the pump amplitude $\frac{\eta}{\gamma}=1.0$ (thick line),$\frac{\eta}{\gamma}=1.4$ (thin line). As the effective lattice height increases,the quantum depletion of the condensate increases.} 
\label{fig:figure_9}
\end{figure}

\begin{figure}[t]
\hspace{-1.5cm}
\includegraphics{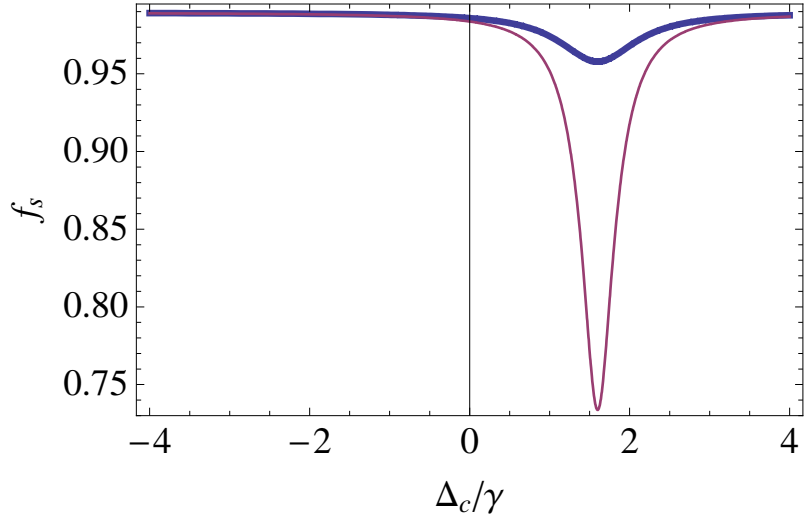} 
\caption{The superfluid fraction as a function of the pump-cavity detuning $\frac{\Delta_{c}}{\gamma}$ for $\frac{JU_{0}}{\gamma}=0.8$, $\frac{E}{\gamma}=-2.0$, $\frac{U}{\gamma}=2.0$, $N=20$, $\frac{\eta}{\gamma}=1.0$ (thick line), $\frac{\eta}{\gamma}=1.2$ (thin line). The superfluid fraction shows a dip at $\Delta_{c}=2n_{0}U_{0}J$, the point where the effective mass shows a maxima. A stronger pump will deplete the condensate more. }
\label{fig:figure_10}
\end{figure}

These expressions give a direct insight into the change of the superfluid fraction as we change the various cavity parameters. A plot (Figure 9) of the superfluid fraction as a function of the barrier height reveals a decrease as the height of the lattice increases. Such an effect i.e quantum depletion of the condensate as the lattice depth increases has been observed recently in an optical lattice without cavity \cite{Xu06}.  The influence of the pump on the superfluid fraction is also shown in the same figure. For a pump amplitude of $\dfrac{\eta}{\gamma}=1$ (thick line) the decrease of the superfluid fraction is slow as the height of the lattice increases. On the other hand when we increase the pump amplitude to $\dfrac{\eta}{\gamma}=1.4$ (thin line), the superfluid fraction depletes much faster for larger lattice heights. As noted earlier, the effective mass shows a maxima at $\Delta_{c}=2n_{0}U_{0}J$. This is accompanied by a decrease in the superfluid fraction and a plot between $f_{s}$ and $\Delta_{c}$ (figure 10) reveals a minima at $\Delta_{c}=2n_{0}U_{0}J$. These results are consistent with earlier work \cite{Maschler05}, where it was found that the fluctuations in the atom number is enhanced (increase in the superfluid fraction) as $\Delta_{c}$ increases (decrease in lattice depth). A direct consequence of the decrease in the superfluid fraction is a decrease in the number fluctuation. This effect can be seen directly by looking at the interference pattern of a BEC released from an optical trap. In contrast to this destructive measurement, one also measure the transmission spectra since various phases of the BEC show qualitatively distinct light scattering \cite{Mekhov07}. The common coupling of all atoms to the same mode introduces cavity-mediated long-range interactions and also the atomic backaction on the field introduces atom-field entanglement \cite{Maschler05}. Consequently, when the atom-number fluctuation changes (which is a direct consequence of a change in the superfluid fraction), the photon-number fluctuations will also change, which will be reflected in the transmission spectra.

Finally, we discuss the experimental feasibility of the proposed experiment. The setup shown in  \figurename{1}, could be part of a ring resonator as described in ref. \cite{Nagorny03}. The ultra-cold atoms are trapped inside an optical standing wave formed by two mutually counter-propagating light waves in a high-Q ring resonator. An external pump can drive the system through one of the side mirrors. A typical problem one encounters experimentally is that the cavity field may not be far detuned from the atomic transition and hence induce spontaneous emission. This will reduce the lifetime of the BEC in the cavity. Yet another experimental fact, that the atoms are coupled to more than one cavity mode may complicate the atom-field dynamics.
The energy of the cavity mode decreases due to the photon loss through the cavity mirrors, which leads to a reduced atom-field coupling. Photon loss can be minimized by using high-Q cavities. Our proposed detection scheme relies crucially on the fact that coherent dynamics dominate over the losses. It is important that the characteristic time-scales of coherent dynamics are significantly faster than those associated with losses (the decay rate ($\gamma$) of state-of-art optical cavities is typically 17 kHz \cite{Klinner06}). Indeed, this fact has become experimentally feasible recently \cite{Klinner06}.  
Experiments with high-Q cavities, which employ standing wave modes tuned close to an atomic transition, rapidly destroy the BEC due to spontaneous emission. Moreover accessing strong atom-field coupling under these circumstances puts severe restriction on the size of the cavities that can be employed. These conditions imply the need for short cavities, which can hold small atomic samples. Recent experiments \cite{Klinner06} have explored a new regime of the interaction between atoms and high-Q cavities: bound atom-cavity systems involving several millions atoms, which operate far from atomic resonances. They showed that in a ring cavity, even at large detunings from the atomic resonance a strong coupling between the atoms and the cavity field can be achieved.

\section{Conclusions}
We have studied the effect of a one dimensional optical lattice in a cavity on the Bloch energy, the effective mass, the Bogoliubov excitation and the superfluid fraction of a Bose-Einstein condensate. The cavity light field develops a photonic band structure due to the strong coupling with the condensate. In the presence of the external pump, the effective mass increases and the Bloch and the Bogoliubov energies are reduced. We have shown that in the cavity, the strong coupling between the atoms and the quantized fields suppresses the superfluidity and hence generates atom-number squeezed state for improved atom-interferometry \cite{Li07}. An additional control over the phase coherence in the superfluid can be achieved by the external pump and the pump-cavity detuning.The effective mass in an optical lattice determines the properties of localized nonlinear excitations such as solitons. Manipulating the amplitude of the external pump and the pump-cavity detuning, we now have a possibility to efficiently control such nonlinear excitations in an optical cavity.

\begin{acknowledgments}
The author is grateful to the Max Planck Institute for Physics of Complex Systems, Dresden, Germany for the hospitality and for providing the facilities for carrying out the present work. The author thanks Anatole Kenfack, Kamal Priya Singh and Sarika Jalan for careful reading of the manuscript and useful discussions.  
\end{acknowledgments}

\end{document}